\documentclass[12pt]{article}

\usepackage[cp1251]{inputenc}
\usepackage{amsmath}
\usepackage{amsfonts}
\usepackage{amssymb}
\usepackage{amsthm}
\usepackage{newlfont}
\usepackage{graphicx}

\date{}
\title{Form factors of the exotic baryons with isospin I=5/2}
\author{ S. M. Gerasyuta $^{1,2}$, M. A. Durnev $^{1}$ \\ {\em $^{1}$ Department of Theoretical Physics }\\ 
\em {St. Petersburg State University, 198904,} \\
\em {St. Petersburg, Russia}\\
\em {$^{2}$ Department of Physics, LTA, 194021,} \\
\em {St. Petersburg, Russia}\\
} 
\scrollmode

\begin{document}
\maketitle

\begin{abstract}
{\small 
\noindent   The electromagnetic form factors of the exotic baryons are calculated in the framework of \; the relativistic \; 
quark \; model \; at \; small \; and \; intermediate \; momentum \; transfer \; $Q^2 \le 1$ GeV$^2$. The charge 
radii of the $E^{+++}$ baryons are determined. 
   \\ 
}
\end{abstract}
{\bf I. Introduction \\}

The consideration of relativistic effects in the composite systems is sufficiently important
when the quark structure of the hadrons is studied [1-10]. The dynamical variables (form factors, scattering
amplitudes) of composite particles can be expressed in terms of the Bethe-Salpeter equations
or quasipotentials. The form factors of the composite particles were considered by a number of 
authors, who have in particular applied a ladder approximation for the Bethe-Salpeter equation [11],
ideas of conformal invariance [12], a number of results was obtained in the framework of 
three-dimensional formalisms [13]. It seems that an application of the dispersion integrals over
the masses of the composite particles may be sufficiently convenient to the description of the
relativistic effects in the composite systems. On the one hand, the dispersion relation 
technique is relativistically invariant one and it is not determined with a consideration of any
distinguished frame of reference. On the other hand, there is no problem of additional 
states arising, because contributions of intermediate states are controlled in the dispersion
relations. The dispersion relation technique allows to determine  the form factors of the composite
particles [14].

    The relativistic generalization of the Faddeev equations  was constructed in the form of 
dispersion relations in the pair energy of two interacting particles and the integral equations  
were obtained for the three-particle amplitudes of $S$-wave baryons: for the octet $J^P=\dfrac {1} {2} ^+$
and the decuplet $J^P=\dfrac {3} {2} ^+$ [15]. The approximate solution of the relativistic three-particle 
problem using the method based on the extraction of the leading singularities 
of the scattering amplitudes about $s_{ik}=4m^2$ was proposed. The three-quark amplitudes given in 
Refs. [15,16] could be used for the calculation of electromagnetic nucleon form factors at small 
and intermediate momentum transfers [17].
    
    In the present paper the computational scheme of the electromagnetic form factors of the exotic baryons 
$(uuuu\bar d)$, consisted of five particles, in the infinite momentum frame is given.

    The nucleon form factors are calculated in Refs. [15, 17] with the help of the dispersion 
relation technique. The proposed approach is generalized to the case of five particles.
     
     Section II is devoted to the calculation of electromagnetic exotic baryon form factors
in the infinite momentum frame. The calculation results of electric form factors of the lowest 
exotic baryons with I=5/2 are given in Section III. The last section is devoted to our discussion
and conclusion.   \\ \\

{\bf II. The calculation of electromagnetic exotic baryon form factors  in the infinite
momentum frame \\}

    Let us consider the electromagnetic form factor of a system of five particles
(an exotic baryon), shown in Fig.1a. The momentum of the exotic baryon is treated to be large: 
 $P_{z} \to \infty$, the momenta $P=k_{1}+k_{2}+k_{3}+k_{4}+k_{5}$ and $P'=P+q$
correspond to the initial and final momenta of the system. Let us assume $P=(P_{0}, \mathbf P_{\perp}=0, P_{z})$   
and $P'=(P'_{0}, \mathbf P'_{\perp}, P'_{z})$. $s$ and $s'$ are the initial and final energy of the system
($P^2=s,\quad P'^2=s'$).
Then we have some conservation laws for the input momenta\\[-30pt] 
{\multlinegap=0pt \begin{multline} \raisetag{30pt} \mathbf k_{1\perp}+\mathbf k_{2\perp}+\mathbf k_{3\perp}+\mathbf k_{4\perp}+\mathbf k_{5\perp}=0 \\
    \shoveleft{P_{z}-k_{1z}-k_{2z}-k_{3z}-k_{4z}-k_{5z}=P_{z}(1-x_{1}-x_{2}-x_{3}-x_{4}-x_{5})=0 } \\
    \shoveleft{P_{0}-k_{10}-k_{20}-k_{30}-k_{40}-k_{50}=P_{z}(1-x_{1}-x_{2}-x_{3}-x_{4}-x_{5})+} \qquad \quad \\ \shoveleft{+
    \dfrac {1} {2P_{z}}\left.\left(s-\dfrac {m^2_{1\perp}} {x_{1}} -  \dfrac {m^2_{2\perp}} {x_{2}} -  
    \dfrac {m^2_{3\perp}} {x_{3}} - \dfrac {m^2_{4\perp}} {x_{4}} -  \dfrac {m^2_{5\perp}} {x_{5}} \right) \right. =0} \\
    \shoveleft{m^2_{i\perp}=m^2+\mathbf k^2_{i\perp}, \quad x_{i}=\dfrac {k_{iz}} {P_{z}}, \quad i=1,2,3,4,5}
     \qquad \qquad \qquad \qquad \qquad  \end{multline} } 
By analogy for the output momenta :\\[-30pt]
    {\multlinegap=0pt \begin{multline}
    \mathbf k'_{1\perp}+\mathbf k_{2\perp}+\mathbf k_{3\perp}+\mathbf k_{4\perp}+\mathbf k_{5\perp}-\mathbf q_{\perp}=0 \\
    \shoveleft{P'_{z}-k'_{1z}-k_{2z}-k_{3z}-k_{4z}-k_{5z}=P_{z}(z-x'_{1}-x_{2}-x_{3}-x_{4}-x_{5})=0} \\
    \shoveleft{P'_{0}-k'_{10}-k_{20}-k_{30}-k_{40}-k_{50}=P_{z}(z-x'_{1}-x_{2}-x_{3}-x_{4}-x_{5})+\qquad \quad}
     \\  \shoveleft{+
   \dfrac {1} {2P_{z}} \left.\left(\dfrac {s'+\mathbf q^2_{\perp}} {z}-\dfrac {m'^2_{1\perp}} {x'_{1}} -  
    \dfrac {m^2_{2\perp}} {x_{2}} -  \dfrac {m^2_{3\perp}} {x_{3}} - \dfrac {m^2_{4\perp}} {x_{4}} -  
    \dfrac {m^2_{5\perp}} {x_{5}} \right) \right. =0} \\
    \shoveleft{x'_{1}=\dfrac {k'_{1z}} {P_{z}}, \quad m'^2_{1\perp}=m^2_{1}+\mathbf k'^2_{1\perp}}
    \qquad \qquad \qquad \qquad \qquad \qquad \qquad \qquad \qquad \end{multline} }

It is introduced in (1) and (2)  $\mathbf q_{\perp} \equiv \mathbf P'_{\perp}$ and 
$z=\dfrac {P'_{z}} {P_{z}}=\dfrac {s'+s-q^2} {2s}$. The form factor of the five-quark system can be obtained 
with the help of the double dispersion integral:
$$F(q^2)= \int\limits_{(m_{1}+m_{2}+m_{3}+m_{4}+m_{5})^2}^{\Lambda_{s}} \dfrac {ds ds'} {4\pi^2} \dfrac {disc_{s} disc_{s'} 
F(s,s',q^2)} {(s-M^2)(s'-M^2)}, \eqno (3) $$
$$ disc_{s} disc_{s'} F(s,s',q^2)=GG'\int d\rho (P,P',k_{1},k_{2},k_{3},k_{4}) \eqno (4) $$
The invariant phase space $d\rho (P,P',k_{1},k_{2},k_{3},k_{4})$, which enters in the double dispersion 
integral, has the form:
 {\multlinegap=0pt \begin{multline}
d\rho (P,P',k_{1},k_{2},k_{3},k_{4})=d\Phi^{(5)}(P,k_{1},k_{2},k_{3},k_{4},k_{5})\times d\Phi^{(5)}(P', k'_{1}, k'_{2},
k_{3}, k'_{4}, k'_{5})\times \\ \shoveleft{\times \prod\limits_{l=2}^{5} (2\pi)^3 2k_{l0} \delta^3(\mathbf k_{l}-\mathbf k'_{l}),}
\qquad \qquad \qquad \qquad \qquad \qquad \qquad  \qquad \qquad \tag{5} \end{multline} }  
where the five-particle phase space is introduced:
$$d\Phi^{(5)}(P,k_{1},k_{2},k_{3},k_{4},k_{5})=(2\pi)^4\delta^4 (P-k_{1}-k_{2}-k_{3}-k_{4}-k_{5})\prod_{l=1}^{5} \dfrac
{d^3 k_{l}} {(2\pi)^{3l}2(k_{l0})^2}$$
After the transformation we have:
{\multlinegap=0pt\begin{multline}d\rho (P,P',k_{1},k_{2},k_{3},k_{4})= \dfrac{1} {2^{10}(2\pi)^{12}} \dfrac {dx_{1}} {x_{1}} d\mathbf k_{1\perp}
\dfrac {dx_{2}} {x_{2}} d\mathbf k_{2\perp} \dfrac {dx_{3}} {x_{3}} d\mathbf k_{3\perp} \dfrac {dx_{4}} {x_{4}} 
d\mathbf k_{4\perp} \times \\ \shoveleft{\times \dfrac {1} {(z-1+x_{1})(1-x_{1}-x_{2}-x_{3}-x_{4})} \times} \\ \shoveleft{\times \delta 
\left.\left(s-\dfrac {m^2_{1\perp}} {x_{1}} -\dfrac {m^2_{2\perp}} {x_{2}}-\dfrac {m^2_{3\perp}} {x_{3}}-\dfrac 
{m^2_{4\perp}} {x_{4}}-\dfrac {m^2_{5\perp}} {1-x_{1}-x_{2}-x_{3}-x_{4}} \right) \right. \times \qquad \quad}
\\ \times \delta \left.\left(\dfrac {s'+\mathbf q^2_{\perp}} {z} -\dfrac {m'^2_{1\perp}} {z-1+x_{1}} -\dfrac 
{m^2_{2\perp}} {x_{2}}-\dfrac {m^2_{3\perp}} {x_{3}}-\dfrac {m^2_{4\perp}} {x_{4}}-\dfrac {m^2_{5\perp}} 
{1-x_{1}-x_{2}-x_{3}-x_{4}} \right) \right. \raisetag{60pt} \tag{6} \end{multline} } \\
For the diquark-spectator (Fig.1b) the invariant phase space takes the more
simplified form:
{\multlinegap=0pt \begin{multline}
d\rho (P,P',k_{1},k_{2},k_{34})= \dfrac{1} {2^{10}(2\pi)^{12}} I_{45} \dfrac {d\mathbf k_{1\perp}} {x_{1}}
\dfrac {d\mathbf k_{2\perp}} {x_{2}} \dfrac {d\mathbf k_{3\perp}} {x_{3}} dx_{1} dx_{2} dx_{3} \dfrac {1} {z-1+x_{1}}
\times \\ \shoveleft{\times \dfrac {1}
{1-x_{1}-x_{2}-x_{3}}  \; \delta \left.\left(s-\dfrac {m^2_{1\perp}} {x_{1}} -\dfrac {m^2_{2\perp}} {x_{2}}
-\dfrac {m^2_{3\perp}} {x_{3}}-\dfrac {m^2_{45\perp}} {1-x_{1}-x_{2}-x_{3}} \right) \right. \times \quad \qquad \quad}
 \\
 \shoveleft{\times \delta \left.\left(\dfrac{s'+\mathbf q^2_{\perp}} {z} -\dfrac {m'^2_{1\perp}} {z-1+x_{1}} -\dfrac 
 {m^2_{2\perp}}{x_{2}}-\dfrac {m^2_{3\perp}} {x_{3}}-\dfrac {m^2_{45\perp}} {1-x_{1}-x_{2}-x_{3}} \right) \right., \quad \quad \; \,}
  \tag{7}  \end{multline} }
where the phase space of the diquark is determined by $I_{45}$  .

    To find the exotic baryon form factor one needs to account the interaction of each 
quark with the external electromagnetic field using the form factor of nonstrange
quarks $f_{q}(q^2)$ [18]. We calculate the $\delta$-functions and obtain for the electromagnetic 
exotic baryon form factor in the case of the normalization $G^{E}(0)=1$:
$$G^{E}(q^2)=\dfrac {F^{E}(q^2)} {F^{E}(0)}=\dfrac {f_{q}(q^2)} {f_{q}(0)} \dfrac {J_{9}(q^2)+J_{12}(q^2)} {J_{9}(0)+
J_{12}(0)},\eqno (8) $$
where:\\ [-30pt]
{\multlinegap=0pt \begin{multline}
J_{9}(q^2)=I_{45}\int\limits_{0}^{\Lambda_{k_{\perp}}} \prod\limits_{i=1}^{3} dk^2_{i\perp} \int\limits_{0}^{1} 
\prod \limits_{i=1}^{3} dx_{i} \int_{0}^{2\pi} \prod \limits_{i=1}^{3} d\phi_{i}\dfrac {1} {x_{1}(1-x_{1})x_{2}(1-x_{2})x_{3}(1-x_{3})}
\times \\
\shoveleft{ \times \dfrac {b\lambda+1} {b+\lambda f} (A^2_{1}+A^2_{4}) \dfrac {\theta(\Lambda_{s}-s) \theta(\Lambda_{s}-s')} {(s-M^2)(s'-M^2)},
 \qquad \qquad \qquad \qquad \qquad \qquad \qquad \qquad \qquad \,}
 \\
\shoveleft{J_{12}(q^2)=\int\limits_{0}^{\Lambda_{k_{\perp}}} \prod \limits_{i=1}^{4} dk^2_{i\perp} \int\limits_{0}^{1} 
\prod \limits_{i=1}^{4} dx_{i} \int_{0}^{2\pi} \prod \limits_{i=1}^{4} d\phi_{i}\dfrac {1} {x_{1}(1-x_{1})x_{2}(1-x_{2})x_{3}(1-x_{3})x_{4} (1-x_{4})}
\times} \\ \shoveleft{\times \dfrac {\tilde b \tilde \lambda+1} {\tilde b+\tilde \lambda \tilde f} A^2_{3} \dfrac {\theta(\Lambda_{s}-
\tilde s)\theta(\Lambda_{s}-\tilde s')} {(\tilde s-M^2)(\tilde s'-M^2)} }
  \qquad \qquad \qquad \qquad \qquad \qquad \qquad \qquad \quad \tag{9} \end{multline} } 
$A_{n}$ ($n$=1,3,4) determine relative contributions of the subamplitudes $BM,\; Mqqq, \; Dqq\bar q$ in the total 
amplitude of the exotic baryon [19], where $B$ and $M$ are the baryon and the meson respectively, while $D$ is the 
diquark. \\
          ($M = 1485$ MeV : $A_{1}=0.3160, A_{3}=0.3393, A_{4}=0.2805;$\\
           $M = 1550$ MeV : $A_{1}=0.2808, A_{3}=0.4209, A_{4}=0.2095$) \\ \\
{\multlinegap=0pt \begin{multline}
 b=x_{1}+ \dfrac {m^2_{1\perp}} {sx_{1}}, \quad f=b^2-\dfrac {4k^2_{1\perp}\cos^2(\phi_{1})} {s}, \quad \lambda= \dfrac
{-b+\sqrt{(b^2-f)\Bigl(1-\Bigl(\dfrac {s} {q^2}\Bigr)f \Bigr) }} {f}, \\
\shoveleft{ s=\dfrac{m^2_{1\perp}} {x_{1}}+\dfrac{m^2_{2\perp}} 
{x_{2}}+\dfrac{m^2_{3\perp}} {x_{3}}+\dfrac{m^2_{45\perp}+k^2_{1\perp}+k^2_{2\perp}+k^2_{3\perp}} {1-x_{1}-x_{2}-x_{3}}+} \\
\shoveleft{+\dfrac{2(\sqrt{k^2_{1\perp}k^2_{2\perp}}\cos (\phi_{2}-\phi_{1})+\sqrt{k^2_{1\perp}k^2_{3\perp}}\cos (\phi_{3}-\phi_{1})+
\sqrt{k^2_{2\perp}k^2_{3\perp}}\cos (\phi_{3}-\phi_{2}))} {1-x_{1}-x_{2}-x_{3}},} \\ \shoveleft{s'=s+q^2(1+2\lambda), 
\qquad \qquad \qquad \qquad \qquad \qquad \qquad \qquad \qquad \qquad \qquad }
 \tag{10}\end{multline} }
{\multlinegap=0pt \begin{multline}
\tilde b=x_{1}+ \dfrac {m^2_{1\perp}} {\tilde sx_{1}}, \quad \tilde f=\tilde b^2-\dfrac {4k^2_{1\perp}\cos^2(\phi_{1})} 
{\tilde s}, \quad \tilde \lambda= \dfrac{-\tilde b+\sqrt{(\tilde b^2-\tilde f)\Bigl (1-\Bigl (\dfrac {\tilde s} {q^2}\Bigr )
\tilde f \Bigr )}} 
{\tilde f}, \\ \shoveleft{\tilde s=\dfrac{m^2_{1\perp}} {x_{1}}+\dfrac{m^2_{2\perp}} {x_{2}}+ \dfrac{m^2_{3\perp}} {x_{3}} + \dfrac 
{m^2_{4\perp}} {x_{4}}+ \dfrac{m^2_{5\perp}+k^2_{1\perp}+k^2_{2\perp}+k^2_{3\perp}+k^2_{4\perp}} {1-x_{1}-x_{2}-x_{3}-x_{4}}
+} \\ \shoveleft{+\dfrac{2(\sqrt{k^2_{1\perp}k^2_{2\perp}}\cos (\phi_{2}-\phi_{1})+\sqrt{k^2_{1\perp}k^2_{3\perp}}\cos (\phi_{3}-\phi_{1})+
\sqrt{k^2_{1\perp}k^2_{4\perp}}\cos (\phi_{4}-\phi_{1}))} {1-x_{1}-x_{2}-x_{3}-x_{4}}+} \\
\shoveleft{ + \dfrac {2(\sqrt{k^2_{2\perp}
k^2_{3\perp}}\cos (\phi_{3}-\phi_{2})+\sqrt{k^2_{2\perp}k^2_{4\perp}}\cos (\phi_{4}-\phi_{2})+\sqrt{k^2_{3\perp}k^2_{4\perp}}
\cos (\phi_{4}-\phi_{3}))} {1-x_{1}-x_{2}-x_{3}-x_{4}},} \\ \shoveleft{\tilde s'=\tilde s+q^2(1+2\tilde \lambda). 
\qquad \qquad \qquad \qquad \qquad \qquad \qquad \qquad \qquad \qquad \qquad} \tag{11}
 \end{multline} } \\[-30pt]

{\bf III. Calculation results}\\

    The electromagnetic exotic baryon form factor  is the sum of two terms(8). The phase 
space of the diquark contributes to the first term $I_{45}=2.036$ GeV$^2$. The vertex functions
$G$ and $G'$ are taken in the middle point of the physical region. The mass of the quarks \, $u, \, d$ 
is equal to $m=0.41$ GeV. The cutoff parameter over the pair energy for the diquarks with \quad  $J^P=1^+ 
\quad \Lambda=20.1$ and the gluon coupling constant $g=0.417$ were obtained in Ref. [19]. It is 
possible to calculate the dimensional cutoff parameters over the total energy and the transvers momentum  
$\Lambda_{s}=33.6$ GeV$^2$, $\Lambda_{k_{\perp}}=0.6724$ GeV$^2$ respectively.
It is necessary to account that the dressed quarks have own
form factors [18]: for $u, \, d$- quarks $f_{q}(q^2)=exp(\alpha_{q}q^2), \, \alpha_{q}=0.33$ GeV$^{-2}$. 
We can use (8) for the numerical calculation of the exotic baryon form factor. It should 
be noted, that the calculation has not any new parameters as compared to the 
calculation of the exotic baryon  mass spectrum [19]. The similar calculation of the proton charge radius 
gives rise to the value $R_{p}= 0.44$ fm, that is almost a factor of two smaller than the experimental value
$R_{p\; \; {\small exp}}  = 0.706$ fm [20]. It is usually  for the quark models with the
one-gluon input interaction [21, 22], when only the presence of the new parameters or the introduction of an
additional interaction allows to achieve a good agreement with the experiment [23, 24].

    The behaviour of the electromagnetic form factor of the exotic baryon $E^{+++}$ with the mass $M=1485$ MeV
is shown in Fig.2. The calculations were carried out
for two exotic baryons with the small masses and the decay widths. We have obtained the orbital angular momentum 
degeneracy [19]. The exotic baryons with the quantum numbers $J^P=\dfrac {1} {2}^+, \dfrac {3} {2}^+, \dfrac {5} {2}^+$ 
with the masses $M$=1485 MeV (the width $\Gamma$=15\;MeV) and $M=1550$ MeV (the width $\Gamma=25$ MeV) are calculated. 
The results turned out to be equal: the charge radius of the $E^{+++}$ baryons $R_{E^{+++}}=0.46$ fm. The charge radius was 
found to be approximately equal to the charge radius of the proton, that  qualitatively corresponds to the 
result of Ref. [25] for the charge radius of the pentaquark $\theta^+(1540)$. It can be concluded that exotic 
baryons are more compact systems than ordinary baryons. The review of experimental results for the $E^{+++}$ 
baryons is given in Ref. [26].
 \\
 \newpage

{\bf IV. Conclusion  }\\

    The method applied in the present work for the study of the exotic baryon form factors  
and based on the transition from the Feynman amplitude to the dispersion integration over the masses of
the composite particles may be extended to the system of $N$ quarks for the multiquark states. One the one hand, the 
calculated result for the form factor of the proton is considerably smaller than the experimental 
value of the proton charge radius. On the other hand, the absence of any new parameters 
introduced in the model for the computation of the exotic baryon form factors is an advantage 
of this method. The qualitative agreement of the results obtained with the calculations in the chiral 
quark-soliton model [25] should be noted.        
\\ \\ 
{\bf Acknowledgments}\\

    The authors would like to thank T. Barnes, S. V. Chekanov, D. I. Diakonov, A. Hosaka and H.-Ch. Kim
for useful discussions. This research was supported by Russian Ministry of Education (Grant 2.1.1.68.26).

\newpage
{\bf References}\\ \\
$1$ \quad F. Gross,  Phys. Lett. B{\bf 140}, 410 (1965).\\
$2$ \quad H. Melosh,  Phys. Rev. D{\bf 9}, 1095 (1974).\\
$3$ \quad G. B. West,  Ann. Phys. (N. Y.) {\bf 74}, 464 (1972).\\
$4$\quad S. J. Brodsky and G. R. Farrar,  Phys. Rev. D{\bf 11}, 1309 (1975).\\
$5$\quad M. V. Terentyev,  Yad. Fiz. {\bf 24}, 207 (1976).\\
$6$\quad V. A. Karmanov,  ZhETF  {\bf 71}, 399 (1976).\\
$7$\quad I. G. Aznauryan and N. L. Ter-Isaakyan,  Yad. Fiz. {\bf 31}, 1680 (1980).\\
$8$\quad A. Donnachie, R. R. Horgen and P. V. Landshoft,  Z. Phys. C{\bf 10}, 71 

(1981).\\
$9$\quad L. L. Frankfurt and M. I. Strikman,  Phys. Rep. C{\bf 76}, 215 (1981).\\
$10$\quad L. A. Kondratyuk and  M. I. Strikman,  Nucl. Phys. A{\bf 426}, 575 (1984).\\
$11$\quad R. N. Faustov,  Ann. Phys. (N. Y.) {\bf 78}, 176 (1973).\\
$12$\quad A. A. Migdal, Phys. Lett. B{\bf 7}, 98 (1971).\\
$13$\quad R. N. Faustov,  Teor. Mat. Fiz. {\bf 3}, 240 (1970).\\
$14$\quad V. V. Anisovich and A. V. Sarantsev,  Yad. Fiz. {\bf 45}, 1479 (1987).\\
$15$\quad S. M. Gerasyuta,   Yad. Fiz. {\bf 55}, 3030 (1992). \\
$16$\quad S. M. Gerasyuta, Z. Phys. C{\bf 60}, 683 (1993).\\
$17$\quad S. M. Gerasyuta, Nuovo Cimento A{\bf 106}, 37 (1993).\\
$18$\quad V. V. Anisovich, S. M. Gerasyuta and A. V. Sarantsev,  Int. J. Mod. 

\quad Phys. A{\bf 6}, 625 (1991).\\
$19$\quad S. M. Gerasyuta and V. I. Kochkin, Phys. Rev. D{\bf 75}, 036005 (2007).\\
$20$\quad M. Gourdin, Phys. Rep. C{\bf 11}, 29 (1974).\\
$21$\quad A. A. Kvitsinsky {\em et. al.}, Yad. Fiz. {\bf 38}, 702 (1986).\\
$22$\quad A. A. Kvitsinsky {\em et. al.}, Fiz. Elem. Chastits At. Yadra {\bf 17}, 267 (1986).\\
$23$\quad F. Cardarelli, E. Pace, G. Salme, and S. Simula, Phys. Lett. B{\bf 357}, 267 

(1995).\\
$24$\quad F. Cardarelli, E. Pace, G. Salme, and S. Simula, nucl-th/9809091.\\
$25$\quad T. Ledwig, H.-Ch. Kim, A. J. Silva, K. Goeke, hep-ph/0603122.\\
$26$\quad A. F. Nilov, Yad. Fiz. {\bf 69}, 918 (2006).\\

%

\newpage
\includegraphics[scale=0.34]{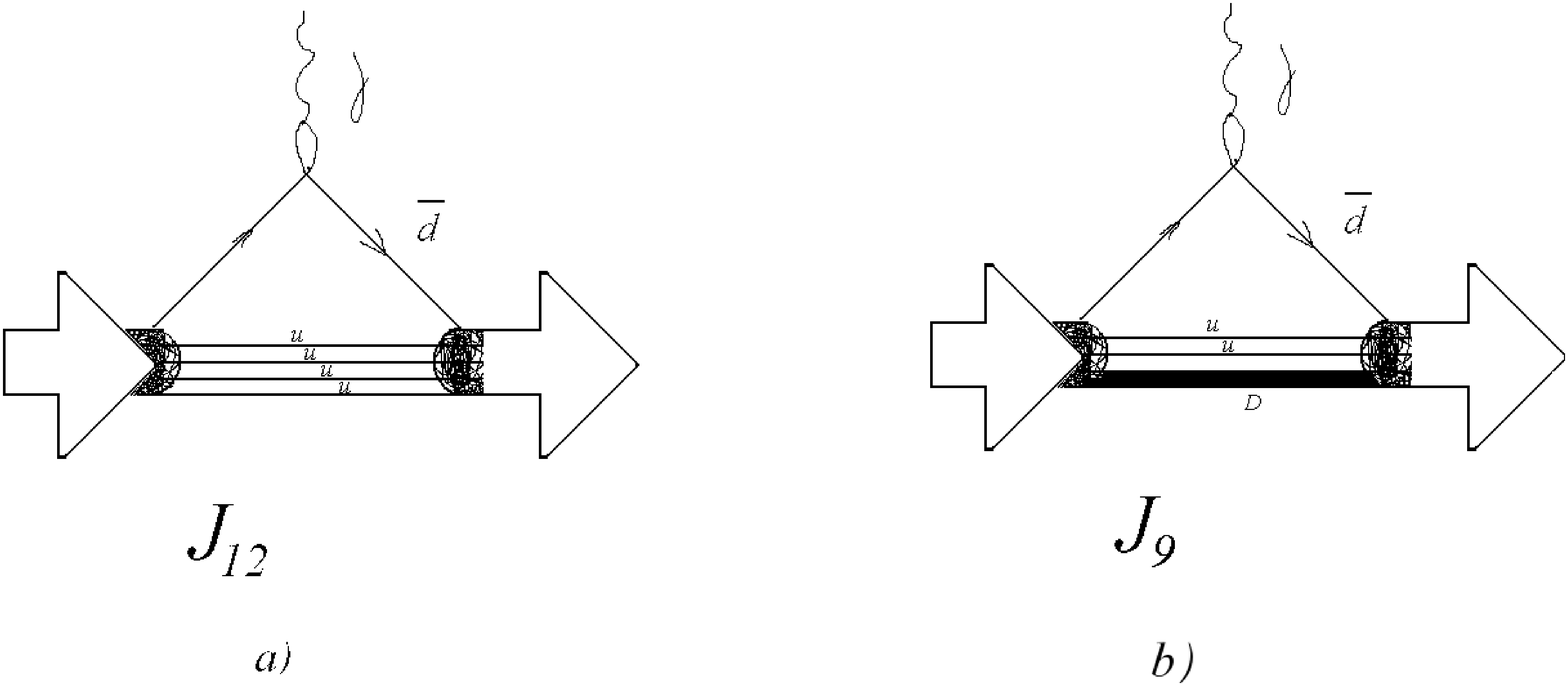}\\
{\bf Fig.1} Triangle diagrams, which determine the form factors of exotic baryons.\\

\newpage
\includegraphics[scale=0.3]{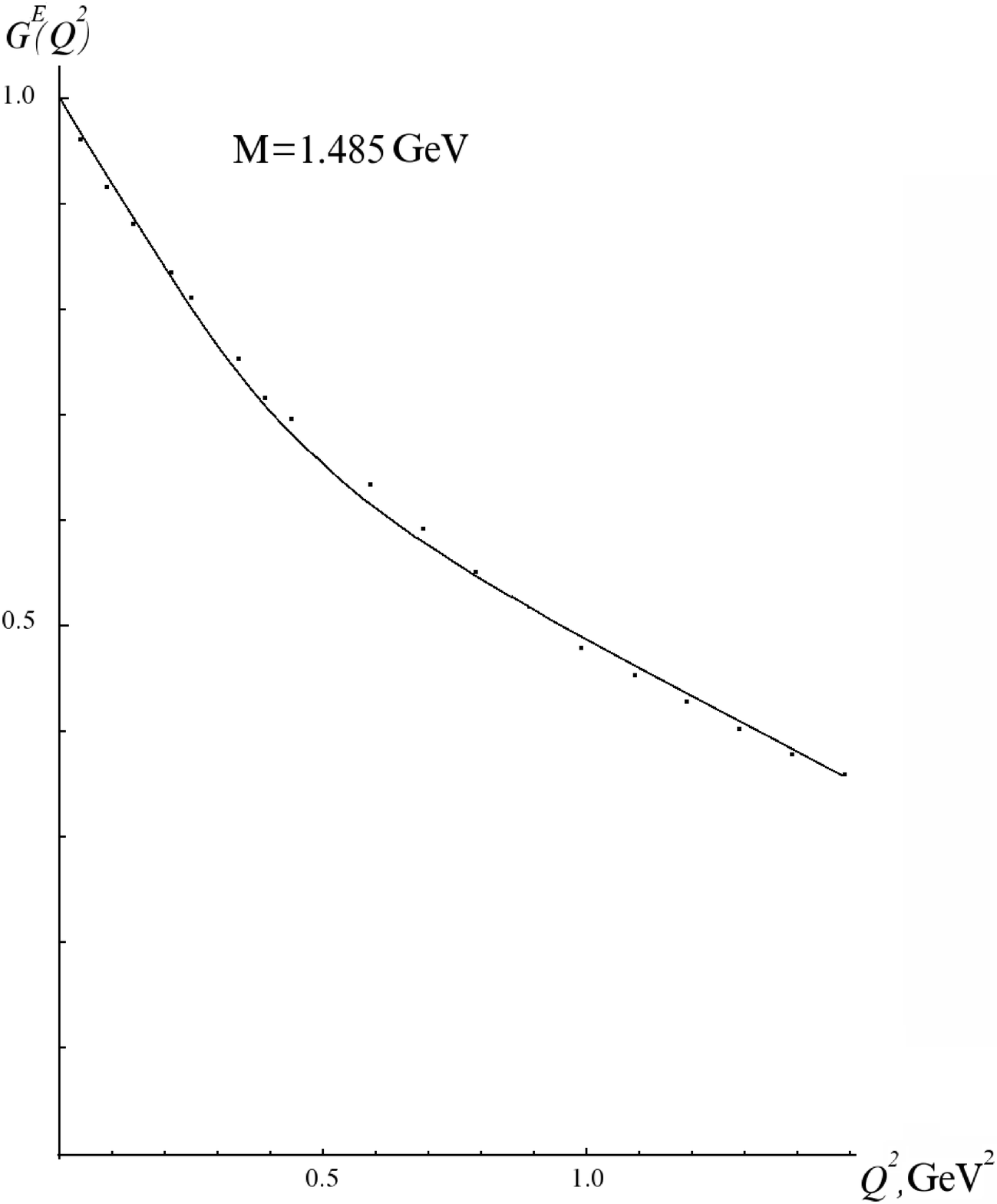}
\\{\bf Fig.2} The electromagnetic form factor of the exotic baryon $E^{+++}$ with mass $M$=1485 MeV and 
decay width $\Gamma$=15 MeV.\\



%

%

\end{document}